# Magnetic and Magneto-Transport Characterization of (Ga,Mn)(Bi,As) Epitaxial Layers

K. Levchenko [1,*], T. Andrearczyk [1], J. Z. Domagala [1], T. Wosinski [1], T. Figielski [1] and J. Sadowski [1,2]

[1] *Institute of Physics, Polish Academy of Sciences, 02-668 Warszawa, Poland*
[2] *MAX-IV Laboratory, Lund University, P.O. Box 118, SE-221 00 Lund, Sweden*

High-quality layers of the (Ga,Mn)(Bi,As) quaternary compound semiconductor have been grown by the low-temperature molecular-beam epitaxy technique. An effect of Bi incorporation into the (Ga,Mn)As ferromagnetic semiconductor and the post-growth annealing treatment of the layers have been investigated through examination of their magnetic and magneto-transport properties. Significant enhancement of the planar Hall effect magnitude upon addition of Bi into the layers is interpreted as a result of increased spin-orbit coupling in the (Ga,Mn)(Bi,As) layers.

*levchenko@ifpan.edu.pl

## 1. Introduction

The ternary III-V semiconductor (Ga,Mn)As has attracted much attention as the model diluted magnetic semiconductor, which combines semiconducting properties with magnetism and offers a basis for developing novel spintronic devices. Substitutional Mn ions in (Ga,Mn)As become ferromagnetically ordered below the Curie temperature owing to interaction with spin-polarized holes. The sensitivity of the magnetic properties, such as the Curie temperature and magnetic anisotropy, to the hole concentration allows for tuning those properties by low-temperature post-growth annealing, photo-excitation or electrostatic gating of the (Ga,Mn)As layers [1, 2]. On the other hand, our recent studies on various types of nanostructures patterned from ferromagnetic (Ga,Mn)As thin layers pointed to their utility for spintronic applications. In particular, nanostructures of the three-arm [3], cross-like [4] and ring-shape [5] geometries, displaying magneto-resistive effects caused by manipulation of magnetic domain walls in the nanostructures, could be applied in a new class of memory cells.

In the present paper we have investigated an impact of Bi incorporation into (Ga,Mn)As layers on their magnetic and magneto-transport properties. Incorporation of a small fraction of Bi, substituting As in GaAs, results in a large decrease in its band gap and the strongly increased spin-orbit coupling, accompanied by a giant separation of the spin-split-off hole band [6, 7]. The increased spin-orbit coupling is especially favourable for spintronic materials where spin precession can be electrically tuned via the Rashba effect.

## 2. Experimental

We have investigated (Ga,Mn)(Bi,As) layers, with 4% and 6% Mn contents and the Bi composition in the range from 0 to 1%, grown by the low-temperature molecular-beam epitaxy (LT-MBE) technique at a temperature of 230°C on semi-insulating (001)-oriented GaAs substrate. After the growth the samples were cleaved into two parts. One part of each sample was subjected to the low-temperature annealing treatment in air at the temperature of 180°C. Annealing at temperatures below the growth temperature can substantially improve magnetic and transport properties of thin (Ga,Mn)As layers due to outdiffusion of charge- and moment-compensating Mn interstitials [8].

High-resolution X-ray diffraction characterization of the investigated layers confirmed their high structural perfection and showed that all of them were grown pseudomorphically on GaAs substrate under compressive misfit strain. The 004 diffraction patterns recorded from the layers showed symmetric (Ga,Mn)(Bi,As) peaks with strong interference fringes, indicating the absence of nonuniformities in the lattice parameters. An addition of a small amount of Bi to the (Ga,Mn)As layers resulted in a distinct increase in their lattice parameter perpendicular to the layer plane and an increase in the in-plane compressive strain.

Magnetic properties of the layers were examined using both the magnetic-field- and temperature-dependent superconducting quantum interference device (SQUID) magnetometry, showing the in-plane ⟨100⟩ easy axes of magnetization in all the layers at low temperatures, typical for (Ga,Mn)As layers grown under compressive misfit strain [4, 9]. Incorporation of Bi into the (Ga,Mn)As layers resulted in decreasing their Curie temperature, $T_C$, and broadening their hysteresis loops. The low-temperature annealing treatment of the (Ga,Mn)(Bi,As) layers resulted in an enhancement of their $T_C$ accompanied with a narrowing their hysteresis loops, i.e. a decrease in the layer coercivity, which is caused mainly by an increase in the hole concentration in annealed layers, as previously observed for the (Ga,Mn)As layers [10, 11].

### 3. Magneto-transport results and discussion

Magneto-transport properties of the layers have been measured in samples of Hall-bar shape supplied with Ohmic contacts to the (Ga,Mn)As and (Ga,Mn)(Bi,As) layers. A picture of representative sample of dimensions of 5 mm in length and 1.5 mm in width, attached to a sample holder, is shown in Fig. 1. Using a low-frequency lock-in technique we measured four-probe longitudinal magneto-resistance (MR) and planar Hall resistance as a function of in-plane magnetic field, $H$, at liquid helium temperatures, in a configuration shown in Fig. 1.

In ferromagnetic materials their resistance depends on the angle $\theta$ between directions of the current, $I$, and the magnetization vector, $M$, resulting in anisotropic magneto-resistance (AMR) and so-called planar Hall effect (PHE) [12]. PHE manifests itself as a spontaneous transverse voltage that develops, owing to the spin-orbit interaction, in response to longitudinal current flow under an in-plane magnetic field or even in the absence of an applied magnetic field. In single domain model the AMR components can be described by expressions [13]:

$$R_{xx} = R_\perp + (R_\parallel - R_\perp)\cos^2\theta, \quad (1)$$

$$R_{xy} = (R_\parallel - R_\perp)\sin\theta\cos\theta, \quad (2)$$

where $R_\perp$ and $R_\parallel$ are the resistances for $I \perp M$ and $I \| M$, respectively. The longitudinal MR is described by Eq. (1) and the transverse resistance, i.e. the planar Hall resistance, is described by Eq. (2). Results obtained for the 10 nm thick (Ga,Mn)As and (Ga,Mn)(Bi,As) layers, of 6% Mn content in the both layers and 1% Bi content in the second one, annealed at 180°C for 50 h, are presented in Figs. 2 and 3. It is worth notice that in the ferromagnetic (Ga,Mn)As, in contradiction to metallic ferromagnets, $(R_\parallel - R_\perp) < 0$, i.e. the resistance is higher when the magnetization is perpendicular to the current with respect to that when they both are parallel [14]. The same also holds in the case of (Ga,Mn)(Bi,As), as confirmed by qualitative similarity of the results obtained for both the (Ga,Mn)As and (Ga,Mn)(Bi,As) layers shown in Figs. 2 and 3.

When sweeping the magnetic field up and down in the range of ±1 kOe, both the MR and PHE resistances vary non-monotonously displaying double hysteresis loops with recurrence points at magnetic fields, which depend on the field orientation. At higher fields, beyond the recurrence points, where the magnetization vector in the layer is forced to be directed along the external magnetic field, longitudinal MR, shown in Fig. 2, exhibits isotropic negative MR typical for the (Ga,Mn)As layers. At low temperatures it results mainly from the magnetic-field induced destruction of quantum interference contribution to the resistivity caused by the effect of weak localization [15, 16]. At lower fields the up-and-down magnetic field sweep causes a rotation of the magnetization vector in the layers by 360° between all the four in-plane ⟨100⟩ directions, which represent equivalent easy axes of magnetization defined by cubic magneto-crystalline anisotropy [14, 17]. As follows from Eqs. (1) and (2), the magnetization vector rotations between the easy magnetization axes results in four transitions of the MR and PHE resistances between their low and high values, observed respectively in Figs. 2 and 3.

Abrupt transitions observed at $H\|I$ configuration, in Figs. 2 and 3a, for both the MR and PHE resistances at magnetic fields of ±105 Oe and ±145 Oe for the (Ga,Mn)As and (Ga,Mn)(Bi,As) layers, respectively, are caused by the magnetization vector rotation by 90° between two perpendicular ⟨100⟩ easy axes of magnetization. The above field values, indicating coercive fields in the layers, confirm that Bi addition into (Ga,Mn)As layers results in an increase in their coercivity. Magnetic field values at which the recurrence points appear, point out the in-plane magnetic anisotropy fields in the investigated layers. At $H\|[\bar{1}10]$ configuration, these anisotropy fields are about 250 Oe and 500 Oe for the (Ga,Mn)As and (Ga,Mn)(Bi,As) layers, respectively, as appears from both the MR and PHE results shown in Figs. 2 and 3a. Under conditions of $H\|[110]$, the much larger magnetic anisotropy fields of about 700 Oe and 900 Oe for the (Ga,Mn)As and (Ga,Mn)(Bi,As) layers, respectively, follow from the PHE results shown in Fig. 3b. All these results indicate easy magnetization axes along the in-plane ⟨100⟩ cubic directions and hard axes along two magnetically non-equivalent in-plane ⟨110⟩ directions, with the [$\bar{1}$10] direction being magnetically easier than the perpendicular [110] one in both the (Ga,Mn)As and (Ga,Mn)(Bi,As) layers. On the other hand, larger magnitudes of both the longitudinal MR changes and the PHE resistances revealed for the (Ga,Mn)(Bi,As) layer, with respect to those for the (Ga,Mn)As layer, indicate for an enhancement of the spin-orbit interaction upon the addition of Bi into the layers.

## 4. Conclusions

Incorporation of a small amount of Bi into the (Ga,Mn)As layers results in decreasing their Curie temperature and increasing coercivity of the layers accompanied with a decrease in their hole concentration. The PHE results reproduce the in-plane magnetic anisotropy and the magnetization reversal in the investigated layers and testify to similarity of the magneto-crystalline anisotropy in both the (Ga,Mn)As and (Ga,Mn)(Bi,As) layers. Stronger spin-orbit coupling in the (Ga,Mn)(Bi,As) layers manifests itself in an enhanced magnitude of the PHE resistance. This quaternary compound proves to be a new diluted magnetic semiconductor displaying properties advantageous for spintronic materials.

## Acknowledgment

This work was supported by the Polish National Science Centre under grant No. 2011/03/B/ST3/02457.

**Figure captions**

Fig. 1. Sample configuration for the magneto-transport measurements. $\theta$ and $\varphi$ denote angles between the current direction and, respectively, the magnetization vector and applied magnetic field.

Fig. 2. Relative sheet longitudinal resistance $R_s$ for the (Ga,Mn)As and (Ga,Mn)(Bi,As) layers measured at the temperature 4.2 K, while sweeping an in-plane magnetic field, parallel to the current, in opposite directions, as indicated by arrows. The curves have been vertically offset for clarity.

Fig. 3. Planar Hall resistance $R_{xy}$ for the (Ga,Mn)As and (Ga,Mn)(Bi,As) layers measured at the temperature 4.2 K, while sweeping an in-plane magnetic field, parallel (a) and perpendicular (b) to the current, in opposite directions, as indicated by arrows. The curves have been vertically offset for clarity.

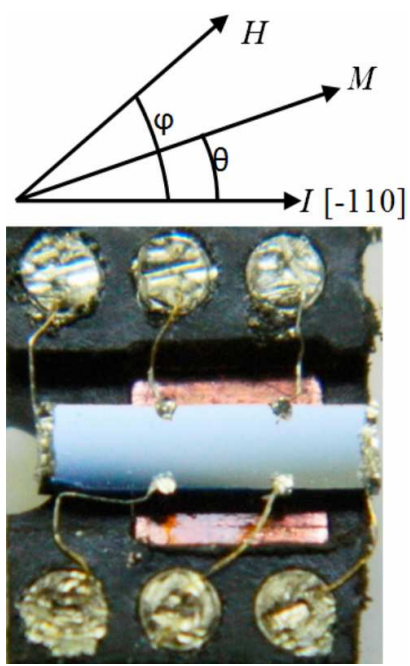

Fig. 1

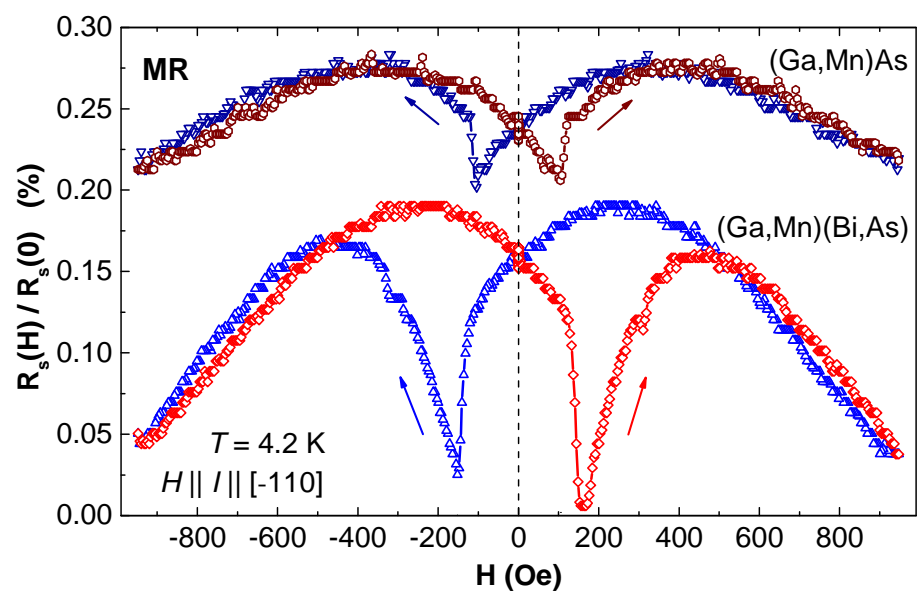

Fig. 2

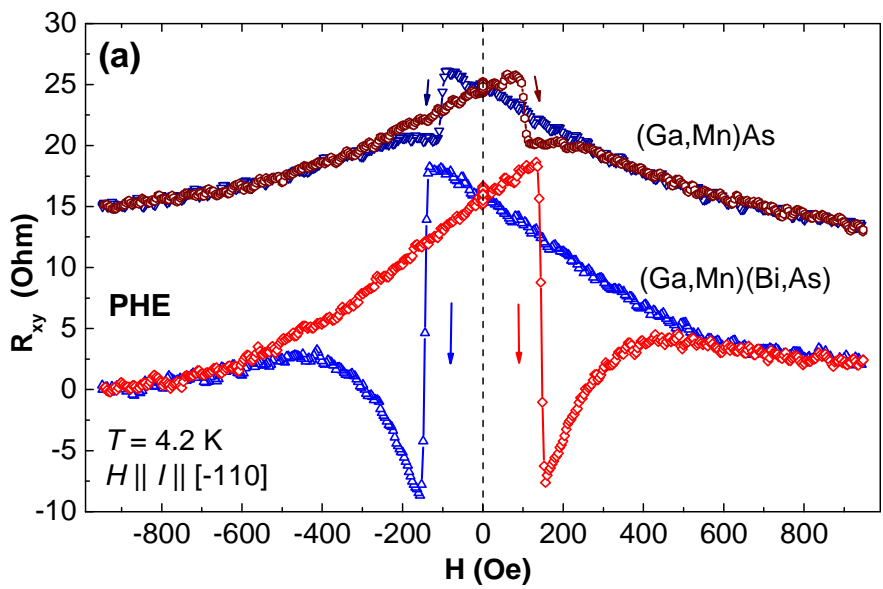

Fig. 3a

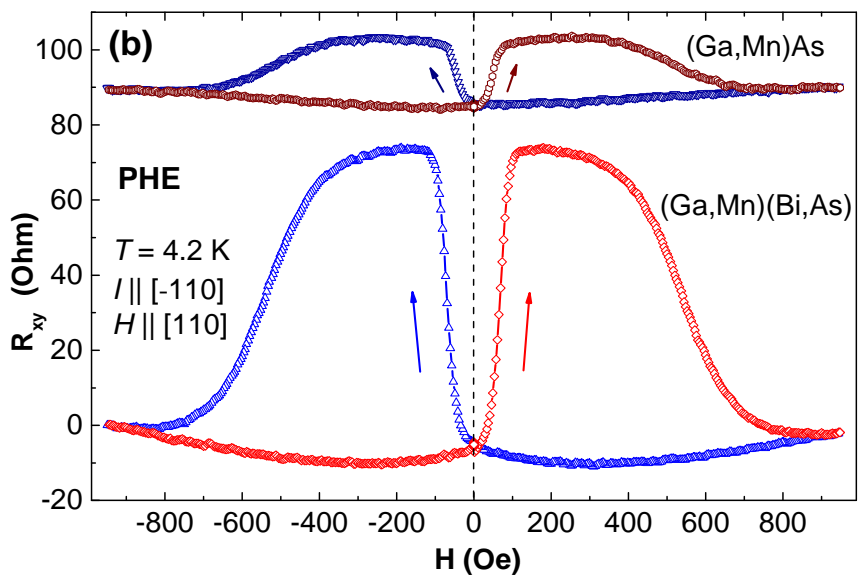

Fig. 3b